\documentclass{jpsj2}

\title{
Conductance and Its Variance 
of Disordered Wires \\ with Symplectic Symmetry in 
the Metallic Regime 
}

\author{
Hiroshi \textsc{Sakai} and Yositake \textsc{Takane}
}
\inst{%
Department of Quantum Matter, Graduate School of Advanced Sciences of
Matter, Hiroshima University, Higashi-Hiroshima 739-8530
}
\recdate{\today}

\abst{%
The conductance of disordered wires with symplectic symmetry is studied 
by a random-matrix approach.
It has been shown that the behavior of the conductance 
in the long-wire limit crucially depends on whether the number of conducting channels is even or odd.
We focus on the metallic regime where the wire length is much shorter
than the localization length, and calculate the ensemble-averaged conductance 
and its variance for both the even- and odd-channel cases. 
We find that the weak-antilocalization correction to
the conductance in the odd-channel case is
equivalent to that in the even-channel case.
Furthermore, we find that the variance dose not depend on whether
the number of channels is even or odd. 
These results indicate that in contrast to the long-wire limit,
clear even-odd differences cannot be observed 
in the metallic regime.
}

\kword{%
disordered quantum wire, symplectic class, metallic regime,
transmission eigenvalue, random-matrix theory
}

\begin{document}
\maketitle

\section{Introduction} 

The quantum transport property of electrons 
in a quasi-one-dimensional disordered wire is
characterized by the symmetries which the system possesses.~\cite{Bee} 
Ordinary disordered wires are classified into either of three
universality classes (orthogonal class, unitary class and symplectic
class) called the standard three. The orthogonal class
consists of systems having both time-reversal symmetry and spin-rotation
invariance, while the unitary class is characterized by the absence of
time-reversal symmetry. The systems having time reversal-symmetry
without spin-rotation invariance belong to the symplectic class.
In the metallic regime where wire length $L$ is much
shorter than the localization length $\xi$, the weak-localization (weak-antilocalization) effect arises in the
orthogonal (symplectic) class, while the effect disappears in the
unitary class. If $L\gg \xi $, the conductance decays exponentially with
increasing $L$ in all the classes.

We focus on the symplectic class in the following. 
The transport property in the symplectic class had been
studied only in the even-channel case.~\cite{Bee,MaCha} However, 
recent theoretical studies reveal that
the behavior in the odd-channel case is very different from that in the
even-channel case.~\cite{AN,NA,AS,YW,YT,YT1,YT2,YT3}
Let $N_{\alpha}$ ($\alpha =\mathrm{even} \ \mathrm{or} \ \mathrm{odd}$) 
be the number of conducting channels. We consider the
dimensionless conductance $g_{\mathrm{even}}$ in the even-channel case
of $N_{\mathrm{even}}=2m$ and $g_{\mathrm{odd}}$ in the odd-channel case
of $N_{\mathrm{odd} }=2m+1$.
When the wire length $L$ is much longer
than the localization length $\xi$, the Anderson localization arises 
in the even-channel case and the averaged conductance behaves as 
\begin{align}
\langle g_{\mathrm{even} } \rangle \to 0 \label{eq:geven0}
\end{align}
with increasing $L$. In the odd-channel case, it has been proved that there exists
one perfectly conducting channel,~\cite{AN,AS} 
and the averaged conductance behaves as 
\begin{align}
\langle g_{\mathrm{odd}} \rangle \to 1 \label{eq:godd1}
\end{align}
in the long-$L$ limit.~\cite{YT,YT2} 
This indicates the absence of Anderson localization. We conclude that in the long-$L$
limit, the behavior of $\langle g_{\mathrm{odd}} \rangle$ is very different from that of
$\langle g_{\mathrm{even}} \rangle$. Therefore, we must separately treat
the odd-channel case and the even-channel case in considering
the symplectic class. However, even-odd differences in
the metallic regime have not been clarified so far.~\cite{comment}

In this paper, we study the conductance of disordered wires with
symplectic symmetry in the metallic regime by a random-matrix approach.
Our attention is focused on differences between the even- and odd-channel cases. We calculate the
ensemble-averaged conductance, as well as its variance,  
by using the scaling approach~\cite{MaS} based on a random-matrix theory.
The variance is defined by 
\begin{align}
\mathrm{Var} g_{\alpha} \equiv \langle g_{\alpha}^{2} \rangle - \langle
g_{\alpha} \rangle^{2}  ,\label{eq:Varg}
\end{align}
where $\alpha =\mathrm{even} \ \mathrm{or} \ \mathrm{odd}$. 
Since one additional channel is perfectly conducting in the
odd-channel case, one may expect 
$\langle g_{\mathrm{odd}} \rangle \stackrel{?}{=} \langle g_{\mathrm{even}} \rangle  +1$.
Contrary to this hypothesis, we obtain 
\begin{align}
\langle g_{\mathrm{odd}} \rangle \sim \langle g_{\mathrm{even} } \rangle \label{eq:evenodd}
\end{align}
in the metallic regime. Indeed, we find that the weak-antilocalization
correction to the averaged conductance does not depend on whether the
number of conducting channels is even or odd.
The reason for this is presented from the
viewpoint of eigenvalue repulsions. 
We also obtain 
\begin{align}
\mathrm{Var} g_{\mathrm{odd} } =\mathrm{Var} g_{\mathrm{even} } . \label{eq:VevenVodd} 
\end{align} 
Our results indicate that the peculiar even-odd difference in the
symplectic universality class appears only in the long-wire regime. 

\section{Conductance and Its Variance} 

The quantum electron transport is described by 
$N_{\alpha} \times N_{\alpha}$ reflection matrices $r$ and
$r^{\prime}$ (reflection from left
to left and from right to right) and transmission matrices $t$ and
$t^{\prime}$ (transmission from left to right and from right to left).
In the symplectic case, which we are interested in, 
the transmission and reflection matrices
satisfy~\cite{AS}
\begin{eqnarray}
{}^tt &=& t^{\prime } , \label{eq:timer}  \\
{}^tr &=& -r \ \ \mathrm{and} \ \ 
{}^tr^{\prime }  = -r^{\prime} \label{eq:noback}.
\end{eqnarray}
Equation (\ref{eq:timer}) holds for arbitrary systems with the time
reversal symmetry, while eq.~(\ref{eq:noback}) characterizes the
peculiarity of the symplectic class.
We can show with eq.~(\ref{eq:noback}) that there exists one channel which is
perfectly transmitting without backscattering only in the odd-channel case.~\cite{AN,AS}
The Hermitian matrix $tt^{\dagger}$ has a 
set of eigenvalues $T_{1} ,T_{2} ,...,T_{m}$ in the even-channel case of
$N_{\mathrm{even}}=2m$. Each eigenvalue is two-fold degenerate due to Kramers degeneracy.
In the odd-channel case of $N_{\mathrm{odd}}=2m+1$, there arises one additional eigenvalue which
corresponds to the perfectly conducting channel. Thus, the dimensionless
conductances are expressed as 
\begin{align}
g_{\mathrm{even}} &= 2\sum_{a=1}^{m} T_{a} ,\label{eq:geven} \\
g_{\mathrm{odd}}  &= 1 + 2\sum_{a=1}^{m} T_{a} .\label{eq:godd} 
\end{align}
The first term in
$g_{\mathrm{odd}}$ represents the contribution from the perfectly conducting channel.
Instead of $T_{1} ,T_{2} ,...,T_{m} $, we employ $\lambda_{1} ,\lambda_{2} ,...,\lambda_{m}$ with
$\lambda_{a} \equiv (1-T_{a} )/T_{a}$ in our following argument.

Our purpose is to obtain the ensemble-averaged conductance and its variance as a
function of the system length $L$. The statistical properties of the conductance are determined by the probability
distribution function $P(\lambda_{1} ,\lambda_{2} ,...,\lambda_{m} ;s)$, 
where $s\equiv L/l$ ($l$: mean free path) is the normalized system length.
It has been shown that the evolution of the probability distribution function
with increasing $L$ is generally described by a Fokker-Planck equation,
which is usually called Dorokhov-Mello-Pereya-Kumar (DMPK) equation in
context of the scaling theory.~\cite{Bee} In the
symplectic class, the DMPK equation depends on whether the number
of the conducting channels is even or odd, and is given by~\cite{MaCha,YT}
\begin{eqnarray}
\frac{\partial P}{\partial s} =
\frac{1}{N_{\alpha} -1}
\sum_{a=1}^{m} \frac{\partial }{\partial \lambda_{a} }
\left(
\lambda_{a} (1+\lambda_{a} ) J_{\alpha} 
\frac{\partial}{\partial \lambda_{a} }
\left(\frac{P}{J_{\alpha} } \right)
\right),\label{eq:DMPK2}
\end{eqnarray}
where $\alpha = \mathrm{even} \ \mathrm{or} \ \mathrm{odd}$.
The even-odd difference is described by $J_{\alpha }$, 
\begin{align}
J_{\mathrm{even} }
&\equiv  
\prod_{b=1}^{m-1} \prod_{a=b+1}^{m} 
|\lambda_{a} -\lambda_{b} |^{4} ,  \label{eq:eJ} \\
J_{\mathrm{odd} }
&\equiv  
\prod_{c=1}^{m}
\lambda_{c}^{2}
\times
\prod_{b=1}^{m-1} \prod_{a=b+1}^{m} 
|\lambda_{a} -\lambda_{b} |^{4} . \label{eq:oJ} 
\end{align}
Note that the factor $\prod_{c=1}^{m} \lambda_{c}^{2}$ in
$J_{\mathrm{odd} }$ represents the eigenvalue repulsion to 
$\lambda_{1} ,\lambda_{2} ,\cdots \lambda_{m}$ arising from the perfectly
conducting eigenvalue.
From eq.(\ref{eq:DMPK2}), we can derive the evolution equation for the
ensemble average $\langle F(\lambda_{1} ,\cdots ,\lambda_{m} ) \rangle$
of an arbitrary function $F(\lambda_{1} ,\cdots ,\lambda_{m} )$,~\cite{MaS}
where
\begin{eqnarray}
\langle F(\lambda_{1} ,\cdots ,\lambda_{m} ) \rangle
= \int \prod_{a=1}^{m} \mathrm{d} \lambda_{a} F(\lambda_{1} ,\cdots ,\lambda_{m} ) P(\lambda_{1} ,\cdots ,\lambda_{m} ;\ s). \label{eq:enave}
\end{eqnarray}
Multiplying both sides of eq.~(\ref{eq:DMPK2}) by $F$ and integrating
over $ \{ \lambda_{a} \} $, we obtain,
\begin{eqnarray}
(N_{\alpha } -1)\frac{\partial \langle F \rangle }{\partial s}
&=&
\Biggl\langle \sum_{a=1}^{m}
\biggl\{
\lambda_{a} (1+\lambda_{a} )
\frac{\partial^{2} F}{\partial \lambda_{a}^{2} }
+ (1+2 \lambda_{a} )\frac{\partial F}{\partial \lambda_{a} }
\biggr\} \nonumber \\ 
&\ & \ \ +2 \sum_{a=1}^{m} \sum_{\substack{b=1 \\b (\neq a)}}^{m}
\frac{\lambda_{a} (1+ \lambda_{a} ) \frac{\partial F}{\partial \lambda_{a} }   
-\lambda_{b} (1+ \lambda_{b} ) \frac{\partial F}{\partial \lambda_{b} }}{
\lambda_{a}  -\lambda_{b} }  \Biggr\rangle  \nonumber \\
&\ & \ \ \ \ \ \ \ \ \ \ 
+ \delta_{\alpha ,\mathrm{odd}} 
\Biggl\langle 
\sum_{a=1}^{m}
\biggl\{
2(1+\lambda_{a} ) \frac{ \partial F}{ \partial \lambda_{a} } 
\biggr\}
\Biggr\rangle .\label{eq:Ffunc}
\end{eqnarray}
In calculating the averaged conductance and its variance, it is
convenient to introduce $\Gamma_{\alpha }$
which is defined as 
\begin{align}
\Gamma_{\alpha} =\sum_{a=1}^{m} T_{a}
=\sum_{a=1}^{m} \frac{1}{1+\lambda_{a} } .\label{eq:Gammadef}  
\end{align}
In terms of $\Gamma_{\alpha}$, the ensemble-averaged dimensionless
conductance for the even- and odd-channel cases is given by
\begin{align}
\langle g_{\alpha } \rangle  
=\delta_{\alpha ,\mathrm{odd} }+2\langle \Gamma_{\alpha} \rangle .\label{eq:g-G} 
\end{align}
The variance of the conductance is also given by
\begin{align}
\mathrm{Var} g_{\alpha} =4
\left(
\langle \Gamma_{\alpha}^{2} \rangle -\langle
\Gamma_{\alpha} \rangle^{2} 
\right) . \label{eq:var1}
\end{align}
Thus, we must calculate $\langle \Gamma_{\alpha}^{p} \rangle$ to obtain the averaged conductance
and its variance.

We calculate $\langle \Gamma_{\alpha}^{p} \rangle$ by using the evolution equation. 
To begin with, let us derive the evolution equation for $\langle \Gamma_{\alpha} \rangle$.
Setting $F=\Gamma_{\alpha }$ in eq.~(\ref{eq:Ffunc}), we obtain  
\begin{align}
(N_{\alpha } -1) \frac{\partial \langle \Gamma_{\alpha} \rangle }{\partial s}
=\bigl\langle -2\Gamma_{\alpha }^{2} 
+\Gamma_{\alpha ,2} -2\Gamma_{\alpha }
\delta_{\alpha ,\mathrm{odd} } \bigr\rangle ,\label{eq:Gamma1} 
\end{align}
where $\Gamma_{\alpha q}$ was defined as $\Gamma_{\alpha q} =\sum_{a=1}^{m} T_{a}^{q} $. 
To solve eq.~(\ref{eq:Gamma1}), we need the quantities 
$\langle \Gamma_{\alpha }^{2} \rangle$ and 
$\langle \Gamma_{\alpha 2} \rangle$.
So we set $F=\langle \Gamma_{\alpha 2} \rangle$
in eq.~(\ref{eq:Ffunc}) and obtain
\begin{align}
(N_{\alpha } -1) \frac{\partial \langle \Gamma_{\alpha 2} \rangle }{\partial s}
&=
\bigl\langle 
4\Gamma_{\alpha }^{2} -8\Gamma_{\alpha } \Gamma_{\alpha 2}
-2(1+2\delta_{\alpha ,\mathrm{odd} } )\Gamma_{\alpha 2} 
+4\Gamma_{\alpha 3}
\bigr\rangle .\label{eq:sample2}
\end{align}
New quantities $\langle \Gamma_{\alpha}^{2} \rangle$, 
$\langle \Gamma_{\alpha} \Gamma_{\alpha 2} \rangle$
and $\langle \Gamma_{\alpha 3} \rangle$ 
appear in the right-hand side of eq.~(\ref{eq:sample2}). 
We will observe that every time we write down an evolution equation,
we find new quantities that have not appeared before on the 
right-hand side. 
This means that we cannot obtain a closed set of coupled equations, and 
thereby cannot obtain an exact solution. To overcome this difficulty, 
we employ the method by Mello and Stone~\cite{MaS}
to obtain an approximate solution for $\langle \Gamma_{\alpha} \rangle$ and
$\langle \Gamma_{\alpha}^{2} \rangle$.
This method is based on a series expansion with respect to $m^{-1}$,
and is applicable to the metallic regime of $s\ll m$.
Thus, we assume in our following argument that $m$ is much larger than unity.

We derive the evolution equation for $\langle \Gamma_{\alpha}^{p} \rangle$,
as well as those for the ensemble-averages 
$\langle \Gamma_{\alpha}^{p} \Gamma_{\alpha 2}\rangle$, 
$\langle \Gamma_{\alpha}^{p} \Gamma_{\alpha 3} \rangle$ and 
$\langle \Gamma_{\alpha}^{p} \Gamma_{\alpha 2}^{2} \rangle$. 
Setting 
$F=\Gamma_{\alpha }^{p} ,\ \Gamma_{\alpha}^{p} \Gamma_{\alpha 2},\ \Gamma_{\alpha}^{p} \Gamma_{\alpha 3}$ 
and $\Gamma_{\alpha}^{p} \Gamma_{\alpha 2}^{2}$ in eq.~(\ref{eq:Ffunc}), 
the evolution equations are given by 
\begin{eqnarray}
(N_{\alpha } -1) \frac{\partial \langle \Gamma_{\alpha}^{p} \rangle }{\partial s}
&=&\bigl\langle -2p \Gamma_{\alpha }^{p+1} +p \Gamma_{\alpha }^{p-1} \Gamma_{\alpha 2} 
\nonumber \\
&\ &
+p(p-1) \Gamma_{\alpha }^{p-2} (\Gamma_{\alpha 2} -\Gamma_{\alpha 3} ) -2p\Gamma_{\alpha }^{p}
\delta_{\alpha ,\mathrm{odd} } \bigr\rangle , \label{eq:Tfunc} \\
(N_{\alpha } -1) \frac{\partial \langle \Gamma_{\alpha}^{p} \Gamma_{\alpha 2}
\rangle }{\partial s}
&=& 
\bigl\langle 4 \Gamma_{\alpha }^{p+2} 
-2(p+4) \Gamma_{\alpha }^{p+1} \Gamma_{\alpha 2}  \nonumber \\
&\ &
-2\{ 1+(p+2)\delta_{\alpha ,\mathrm{odd} }
\} \Gamma_{\alpha }^{p} \Gamma_{\alpha 2} 
+4\Gamma_{\alpha }^{p} \Gamma_{\alpha 3}  \nonumber \\
&\ &
+p\Gamma_{\alpha}^{p-1} \Gamma_{\alpha 2}^{2}
+4p\Gamma_{\alpha}^{p-1} (\Gamma_{\alpha 3} - \Gamma_{\alpha 4} ) \nonumber \\
&\ &
+p(p-1)\Gamma_{\alpha }^{p-2} (\Gamma_{\alpha 2}^{2} -\Gamma_{\alpha 2} \Gamma_{\alpha 3} )
\bigr\rangle , \label{eq:T2func}  \\
(N_{\alpha } -1) \frac{\partial \langle \Gamma_{\alpha}^{p} \Gamma_{\alpha 3}
\rangle }{\partial s}
&=&
\bigl\langle 
-2(p+6)
\Gamma_{\alpha }^{p+1} \Gamma_{\alpha 3}
+12 \Gamma_{\alpha }^{p+1} \Gamma_{\alpha 2}
-6 \Gamma_{\alpha }^{p} \Gamma_{\alpha 2}^{2} \nonumber \\
&\ &
-2\{ 3+(p+3)\delta_{\alpha ,\mathrm{odd} }
\} 
\Gamma_{\alpha }^{p} \Gamma_{\alpha 3} 
+9\Gamma_{\alpha }^{p} \Gamma_{\alpha 4} \nonumber \\
&\ & +p\Gamma_{\alpha}^{p-1} \Gamma_{\alpha 2} \Gamma_{\alpha 3}
+6p\Gamma_{\alpha }^{p-1}
(\Gamma_{\alpha 4} -\Gamma_{\alpha 5} ) \nonumber \\
&\ &
+p(p-1)\Gamma_{\alpha }^{p-2}
(\Gamma_{\alpha 2} \Gamma_{\alpha 3} - \Gamma_{\alpha 3}^{2} )
\bigr\rangle , \label{eq:T3func} \\
(N_{\alpha } -1) \frac{\partial \langle \Gamma_{\alpha}^{p} \Gamma_{\alpha 2}^{2}
\rangle }{\partial s}
&=&
\bigl\langle 
-2(p+8)
\Gamma_{\alpha }^{p+1} \Gamma_{\alpha 2}^{2}
+8 \Gamma_{\alpha }^{p+2} \Gamma_{\alpha 2}
+p \Gamma_{\alpha }^{p-1} \Gamma_{\alpha 2}^{3} \nonumber \\
&\ &
-2\{
2+(p+4) \delta_{\alpha ,\mathrm{odd} } 
\} \Gamma_{\alpha }^{p} \Gamma_{\alpha 2}^{2}
+8\Gamma_{\alpha}^{p} \Gamma_{\alpha 2} \Gamma_{\alpha 3} \nonumber \\ 
&\ &
+p(p-1)\Gamma_{\alpha }^{p-2} (\Gamma_{\alpha 2} - \Gamma_{\alpha 3} ) \Gamma_{\alpha 2}^{2}
+8\Gamma_{\alpha}^{p} (\Gamma_{\alpha 4} -\Gamma_{\alpha 5 } )  \nonumber \\
&\ &
+8p\Gamma_{\alpha }^{p-1} \Gamma_{\alpha 2} (\Gamma_{\alpha 3} -\Gamma_{\alpha 4} )
\bigr\rangle .\label{eq:T22func} 
\end{eqnarray}
We seek the solution for the above equations as a series in decreasing
powers of $m$. To do so, we expand the averages as
\begin{eqnarray} 
\langle \Gamma_{\alpha }^{p} \rangle
&=& 
m^{p} f_{\alpha , p,0} (s) +m^{p-1} f_{\alpha ,p,1} (s)
+m^{p-2} f_{\alpha ,p,2} (s) + \cdots ,\label{eq:fseri} \\
\langle \Gamma_{\alpha }^{p} \Gamma_{\alpha 2} \rangle
&=& 
m^{p+1} j_{\alpha , p+1,0} (s) +m^{p} j_{\alpha ,p+1,1} (s)
+m^{p-1} j_{\alpha ,p+1,2} (s) + \cdots ,\label{eq:gseri} \\
\langle \Gamma_{\alpha }^{p} \Gamma_{\alpha 3} \rangle
&=& 
m^{p+1} k_{\alpha , p+1,0} (s) +m^{p} k_{\alpha ,p+1,1} (s)
+m^{p-1} k_{\alpha ,p+1,2} (s) + \cdots ,\label{eq:hseri} \\
\langle \Gamma_{\alpha }^{p} \Gamma_{\alpha 2}^{2} \rangle
&=& 
m^{p+2} l_{\alpha , p+2,0} (s) +m^{p+1} l_{\alpha ,p+2,1} (s)
+m^{p} l_{\alpha ,p+2,2} (s) + \cdots ,\label{eq:lseri} 
\end{eqnarray}
where $f_{\alpha ,p,n} (s)$, $j_{\alpha ,p,n} (s)$, $k_{\alpha ,p,n} (s)$ and $l_{\alpha ,p,n} (s)$ 
are functions of $s$ and satisfy the initial conditions
$f_{\alpha ,p,n} (0)=j_{\alpha ,p,n} (0)=k_{\alpha ,p,n} (0)=l_{\alpha
,p,n} (0)=\delta_{n0}$.

To get the coefficients $f_{\alpha ,p,n} (s)$, $j_{\alpha ,p,n} (s)$,
$k_{\alpha ,p,n} (s)$ and $l_{\alpha ,p,n} (s)$, we substitute the above
expansion series into eqs.~(\ref{eq:Tfunc})-(\ref{eq:T22func}) and equate the coefficients of
the various powers of $m$. 
We obtain the following differential equations, 
\begin{align}
&\ f_{\alpha ,p,0}^{\prime } (s) + pf_{\alpha ,p+1,0} (s) =0 ,\label{eq:fequ} \\
&\ f_{\alpha ,p,1}^{\prime } (s) + pf_{\alpha ,p+1,1} (s) =\frac{1}{2} 
\left[
pj_{\alpha ,p,0} (s) + f_{\alpha ,p,0}^{\prime } (s) \delta_{\alpha ,\mathrm{even} } -2p
f_{\alpha ,p,0} (s) \delta_{\alpha , \mathrm{odd} }
\right] , \label{eq:fdequ} \\
&\ j_{\alpha ,p,0}^{\prime } (s)
+(p+3)j_{\alpha ,p+1,0} (s) =2f_{\alpha ,p+1,0} (s) ,\label{eq:gequ} \\
&\ \ \ \ \ \ \ \ \ \ \ \ \ \ \ \ \ \ \ \ \ \ \ \ \ \ \vdots \nonumber
\end{align}
for the coefficients.
\ \ To derive $f_{\alpha ,p,2} (s)$, we need four equations for  
$l_{\alpha , p,0} (s)$, $k_{\alpha ,p,0}(s)$, $j_{\alpha ,p,1} (s)$ and 
$f_{\alpha ,p,2} (s)$ other than eqs.~(\ref{eq:fequ})-(\ref{eq:gequ}). 
We briefly describe the method to solve the equation for
$f_{\alpha ,p,1} (s)$. Before deriving $f_{\alpha ,p,1} (s)$, we have to get
$f_{\alpha ,p,0} (s)$ and $j_{\alpha ,p,0} (s)$.
By considering the series expansion of $f_{\alpha ,p,0} (s)$ at $s=0$
on the basis of eq.~(\ref{eq:fequ}), we expect that  
\begin{align} 
f_{\alpha ,p,0} (s)=\frac{1}{(1+s)^{p}} . \label{eq:f0}
\end{align}
We can easily show that eq.~(\ref{eq:f0}) satisfies eq.~(\ref{eq:fequ}) for arbitrary
$s$. We observe that $f_{\alpha ,p,0} (s)$ does not depend on whether the
number of channels is even or odd. Next, we derive $j_{\alpha ,p,0} (s)$ in eq.~(\ref{eq:gequ}).
Since the resulting power series for $j_{\alpha ,p,0} (s)$ at $s=0$ is
not easily identifiable, it is difficult to derive $j_{\alpha ,p,0} (s)$
in the above manner. We assume that the solution of eq.~(\ref{eq:gequ}) takes the form of
\begin{align}
j_{\alpha ,p,0} (s) 
=\frac{1}{(1+s)^{p+n} } (a_{r} s^{r} +a_{r-1} s^{r-1} +\cdots +1),
\label{eq:katei}
\end{align}
where $n$ and $r$ are integers to be determined and  $\{a_{r} \}$ are
coefficients which are independent of $p$. From eq.~(\ref{eq:katei}), 
we can obtain the relation, $j_{\alpha ,p+1,0} (s) =(1+s)^{-1} j_{\alpha ,p,0} (s)$.
Substituting this into eq.~(\ref{eq:gequ}), we obtain a solvable
differential equation. The solution is
\begin{align}
j_{\alpha ,p,0} (s) =\frac{1}{3(1+s)^{p+3} } (2s^{3} +6s^{2} +6s +3) .\label{eq:gsol}
\end{align}
Substituting eqs.~(\ref{eq:f0}) and (\ref{eq:gsol}) and the derivative of eq.~(\ref{eq:f0})
into eq.~(\ref{eq:fdequ}), we obtain
\begin{align}
f_{\alpha ,p,1}^{\prime } (s) +pf_{\alpha ,p+1,1} (s)
&= \frac{p}{6(1+s)^{p+3} } 
\left[
2(1-3\delta_{\alpha ,\mathrm{odd} } ) s^{3} \right. +3(1
-5 \delta_{\alpha ,\mathrm{odd} } ) s^{2} \nonumber \\
&\ \ \ \ \ \ \left.  -12s\delta_{\alpha ,\mathrm{odd} } -3\delta_{\alpha ,\mathrm{odd} }
\right] .   \label{eq:f1equ}
\end{align}
We assume that the solution for $f_{\alpha ,p,1} (s)$  
takes the form of $f_{\alpha ,p,1} (s) =p\cdot \chi_{\alpha ,p} (s) +\eta_{\alpha ,p} (s)$,
where $\chi_{\alpha ,p} (s)$ and $\eta_{\alpha ,p} (s)$ take the 
form similar to eq.~(\ref{eq:katei}).
By means of the above assumption, we substitute
$f_{\alpha ,p,1}^{\prime } (s)$ and $f_{\alpha ,p+1,1} (s)$ into 
the left-hand side of eq.~(\ref{eq:f1equ}) 
and obtain solvable differential equations for $\chi_{\alpha ,p} (s)$ and
$\eta_{\alpha ,p} (s)$. We solve the resulting equations and obtain
\begin{align}
f_{\alpha ,p,1} (s)
=\frac{p}{6(1+s)^{p+2} }
\left(
(1-3\delta_{\alpha ,\mathrm{odd} })s^{3}
-6s^{2} \delta_{\alpha ,\mathrm{odd} } -3s\delta_{\alpha ,\mathrm{odd} }
\right) .
\label{eq:f1kai} %f1(s)
\end{align}
Note that $f_{\alpha ,p,1} (s)$ depends on whether the number of
channels are even or odd. The same procedure can be applied to the equations for
$l_{\alpha ,p,0} (s)$, $k_{\alpha ,p,0} (s)$, $j_{\alpha ,p,1} (s)$ and $f_{\alpha ,p,2} (s)$.
The result for $f_{\alpha ,p,2} (s)$ is
\begin{align} 
f_{\alpha ,p,2} (s)
&= \frac{p}{360(1+s)^{p+4}}
\Bigl[
\{ 
11p-9 +(15p-15) \delta_{\alpha ,\mathrm{odd} }
\} s^{6}
\nonumber \\
& \ \ \ \ \ \ \ \ \ \ \ \ \ \ \ 
+
\{
36p-24 
+(120p-120) \delta_{\alpha ,\mathrm{odd} }
\} s^{5} 
\nonumber \\
& \ \ \ \ \ \ \ \ \ \ \ \ \ \ \ 
+
\{
90p-60 +(240p-240) \delta_{\alpha ,\mathrm{odd} }
\} s^{4}
\nonumber \\
& \ \ \ \ \ \ \ \ \ \ \ \ \ \ \ 
+
\{
120p-150+(180p-180)\delta_{\alpha ,\mathrm{odd} }
\} s^{3}
\nonumber \\
& \ \ \ \ \ \ \ \ \ \ \ \ \ \ \ 
+
\{
90p-90+(45p-45)\delta_{\alpha ,\mathrm{odd} }
\} s^{2}
\Bigr]. \label{eq:f2kai} %f2(s)
\end{align}
\ \ Substituting eqs.~(\ref{eq:f0}), (\ref{eq:f1kai}) and (\ref{eq:f2kai})
into eq.~(\ref{eq:fseri}), we finally obtain,
\begin{align} 
\langle \Gamma_{\alpha }^{p} \rangle
&=
m^{p} \frac{1}{(1+s)^{p} } 
+m^{p-1}
\frac{p}{6(1+s)^{p+2} }
\left[
s^{3} -(3s^{3} +6s^{2} +3s)\delta_{\alpha ,\mathrm{odd} }
\right] \nonumber \\
& \ \ \ \ \ \ \ \ \ \ \ 
+ m^{p-2} \frac{p}{360(1+s)^{p+4}}
\Bigl[
\{ 11p-9+(15p-15)\delta_{\alpha ,\mathrm{odd} } \} s^{6}
\nonumber \\
& \ \ \ \ \ \ \ \ \ \ \ \ \ \ \ \ \ \ \ \ \ \ \ \ \ \ \ \ \ \ \ \ \ \ 
+
\{ 36p-24+(120p-120)\delta_{\alpha ,\mathrm{odd} } \} s^{5}
\nonumber \\
& \ \ \ \ \ \ \ \ \ \ \ \ \ \ \ \ \ \ \ \ \ \ \ \ \ \ \ \ \ \ \ \ \ \ 
+
\{ 90p-60+(240p-240)\delta_{\alpha ,\mathrm{odd} }\} s^{4}
\nonumber \\
& \ \ \ \ \ \ \ \ \ \ \ \ \ \ \ \ \ \ \ \ \ \ \ \ \ \ \ \ \ \ \ \ \ \ 
+
\{ 120p-150+(180p-180)\delta_{\alpha ,\mathrm{odd} }\} s^{3}
\nonumber \\
& \ \ \ \ \ \ \ \ \ \ \ \ \ \ \ \ \ \ \ \ \ \ \ \ \ \ \ \ \ \ \ \ \ \ 
+
\{ 90p-90+(45p-45)\delta_{\alpha ,\mathrm{odd} }\} s^{2}
\Bigr]
+ \cdots .\label{eq:Gamma} %Gamma?IE1/21/4Ž°
\end{align}

From eqs.~(\ref{eq:g-G}) and (\ref{eq:Gamma}), we obtain
the ensemble-averaged dimensionless conductance 
$\langle g_{\alpha} \rangle$. 
Setting $p=1$ in eq.~(\ref{eq:Gamma}) and substituting the resulting
expression for $\langle \Gamma_{\alpha} \rangle$ into eq.~(\ref{eq:g-G}),
we obtain
\begin{align}
\langle g_{\mathrm{odd}} \rangle
&=
(2m+1) \frac{1}{1+s} 
 +\frac{s^{3} }{3(1+s)^{3} } \nonumber \\
& \ \ \ \ \ \ \ \ \ \ \ 
+m^{-1} \frac{1}{90(1+s)^{5} } (s^{6} +6s^{5} +15s^{4} -15s^{3} )
+O \left( m^{-2} \right) ,\label{eq:go} \\ %godd 
\langle g_{\mathrm{even}} \rangle
&=
2m \frac{1}{1+s} 
+\frac{s^{3} }{3(1+s)^{3} } \nonumber \\
& \ \ \ \ \ \ \ \ \ \ \ 
+m^{-1} \frac{1}{90(1+s)^{5} } (s^{6} +6s^{5} +15s^{4} -15s^{3} )
+O \left( m^{-2} \right) . \label{eq:ge}    %geven
\end{align}

Now, we discuss the behavior of the 
averaged conductance in the metallic regime of $s\ll m$. 
Note that the even-odd difference appears
only in the first term of eqs.~(\ref{eq:go}) and (\ref{eq:ge}), 
which represents the classical contribution. 
The second term in $\langle g_{\alpha } \rangle$ represents the  
weak-antilocalization correction.
We observe that the weak-antilocalization
correction to $\langle g_{\mathrm{odd}} \rangle$ is equivalent to that to            
$\langle g_{\mathrm{even}} \rangle$. Note that the perfectly conducting
channel is present only in the odd-channel case, and that if we neglect
this special channel, the total number of conducting channels 
in the odd-channel case is equivalent to that in the even-channel case. 
Thus, one may expect 
\begin{align}
\langle g_{\mathrm{odd}}
\rangle \stackrel{?}{=} \langle g_{\mathrm{even}} \rangle  +1 .\label{eq:gyosou}
\end{align}
However, contrary to this hypothesis, our result indicates that
$\langle g_{\mathrm{odd}} \rangle \sim \langle g_{\mathrm{even}} \rangle$.
Indeed, the difference $\langle g_{\mathrm{odd} } \rangle -\langle
g_{\mathrm{even} } \rangle$ 
is much smaller than unity except for the ballistic limit of $s\to 0$.
To explain this, we must consider the repulsion acting between
transmission eigenvalues of $tt^{\dagger}$.~\cite{YT} In the odd-channel case, one transmission
eigenvalue is equal to unity due to the presence of the perfectly
conducting channel. Although it positively contributes to 
$\langle g_{\mathrm{odd} } \rangle$, the other eigenvalues are reduced 
due to the repulsion from it. 
To observe this, we set $p=1$ in eq.~(\ref{eq:Gamma}) and present the expression
for $\Gamma_{\alpha}$,
\begin{align}
\langle \Gamma_{\alpha } \rangle
=
m \frac{1}{(1+s)} 
+\frac{s^{3}}{6(1+s)^{3} } 
-\frac{s}{2(1+s)} \delta_{\alpha ,\mathrm{odd} }
+ O\left(  m^{-1} \right) . \label{eq:Gammap1}
\end{align}
The third term in eq.~(\ref{eq:Gammap1})
represents this reduction. Our result indicates that the contribution to 
$\langle g_{\mathrm{odd}} \rangle$ from the perfectly conducting channel
is almost canceled by the reduction of the contribution from the other channels.

Next, we obtain the variance $\mathrm{Var} g_{\mathrm{\alpha} }$ 
for the even- and odd-channel cases. From eq.~(\ref{eq:Gamma}), we can calculate 
$\langle \Gamma_{\alpha}^{2} \rangle$ and $\langle \Gamma_{\alpha} \rangle^{2}$.
Substituting these quantities into eq.~(\ref{eq:var1}), we obtain
\begin{align}
\mathrm{Var} g_{\mathrm{odd}} &= \mathrm{Var} g_{\mathrm{even}} \nonumber \\
&= \frac{2}{15(1+s)^{6}} (s^{6} +6s^{5} +15s^{4} +20s^{3} +15s^{2} )
+O \left( m^{-1} \right) .\label{eq:var}
\end{align}
The variance does not depend on whether the total number of channels is even or odd.

\section{Summary}

We have studied the conductance and its variance of disordered wires 
with symplectic symmetry in the metallic regime. By using the scaling
approach based on a random-matrix theory, we have 
derived the evolution equations for various quantities, such as 
$\langle \Gamma_{\alpha}^{p} \rangle$ and
$\langle \Gamma_{\alpha}^{p} \Gamma_{\alpha 2} \rangle$,
from which we obtain $\langle g_{\alpha} \rangle$ and $\mathrm{Var} g_{\mathrm{\alpha}}$ 
($\alpha =\mathrm{even} \ \mathrm{or} \ \mathrm{odd}$).
We have solved the resulting equations by using the series expansion with respect
to $s/m$.~\cite{MaS}
It is shown that $\langle g_{\mathrm{odd} } \rangle  \sim \langle g_{\mathrm{even} } \rangle$
and $\mathrm{Var} g_{\mathrm{odd}} = \mathrm{Var} g_{\mathrm{even}}$ in
the metallic regime. 
This means that clear even-odd differences do not appear in the transport properties in the metallic regime. 
We conclude that the even-odd difference can be observed only in the long-wire limit.

This work was supported in part by a Grant-in-Aid for Scientific
Research (C) from the Japan Society for the Promotion of Science.

\end{document}